\documentclass[%
 reprint,
superscriptaddress,
 amsmath,amssymb,
 aps,
prapplied, floatfix, longbibliography ]{revtex4-2}
\usepackage{graphicx}
\usepackage{dcolumn}
\usepackage{bm}
\newcommand{\modR}[1]{                 {#1}}
\usepackage[colorlinks,linkcolor=blue,anchorcolor=blue,citecolor=blue,urlcolor=blue]{hyperref}
\begin{document}

\title{Ferroelectricity in oxygen-terminated 2D carbides of lanthanide elements}

\author{Lin~Han}
\affiliation{School of Physics,
             Southeast University,
             Nanjing, 211189, PRC}

\author{Wencong~Sun}
\affiliation{School of Physics,
             Southeast University,
             Nanjing, 211189, PRC}

\author{Pingwei~Liu}
\affiliation{School of Physics,
             Southeast University,
             Nanjing, 211189, PRC}

\author{Xianqing~Lin}
\affiliation{College of Science,
             Zhejiang University of Technology,
             Hangzhou, 310023, PRC}

\author{Dan~Liu}
\email      {liudan2@seu.edu.cn}%
\affiliation{School of Physics,
             Southeast University,
             Nanjing, 211189, PRC}

\author{David Tom\'{a}nek}
\email      {tomanek@msu.edu}%
\affiliation{Physics and Astronomy Department,
             Michigan State University,
             East Lansing, Michigan 48824, USA}
\affiliation{Department of Physics,
             University of Johannesburg,
             Johannesburg, South Africa}

\date{\today}

\begin{abstract}
We investigate the properties of oxygen-functionalized carbides of
lanthanide elements with the composition M$_2$CO$_2$ (M=Gd, Tb,
Dy) that form two-dimensional (2D) structures. Our {\em ab initio}
calculations reveal that oxygen termination turns M$_2$C
monolayers into \modR{semiconductors} with two dynamically stable
phases. Of these, the energetically favored $\alpha-$phase becomes
ferroelectric, whereas the $\beta-$phase turns anti-ferroelectric.
Applying in-plane biaxial strain may transform one phase into the
other, changes the ferroelectric polarization of the
$\alpha-$phase in a linear fashion, and modifies the size and
nature of the fundamental band gap from direct to indirect. The
structure with a direct band gap exhibits in-plane isotropic
electronic and optical properties. This previously unexplored
class of systems also exhibits excellent photon absorption in the
ultraviolet range.
\end{abstract}




\maketitle
\renewcommand\thesubsection{\arabic{subsection}}




\section{Introduction}

Since the discovery of ferroelectricity in Rochelle salt in
1920s~\cite{{Valasek1920},{Valasek1921}}, ferroelectric materials
have captured ample attention of the scientific community due to
their complex behavior and unique applicability in electronic
devices. The interest in 2D materials, ignited by the successful
exfoliation of graphene, led to the exploration of
ferroelectricity not only in three-dimensional (3D) materials, but
also in lower dimensions. Early studies focused on intrinsic
ferroelectric behavior in 2D compounds including
In$_2$Se$_3$~\cite{{Zhu2017},{Lai17}},
CuInP$_2$S$_6$~\cite{{Kalinin15},{Fucai16}}, and in elemental
materials such as 2D $\alpha-$ or $\delta$-Se~\cite{DT291}. Among
the plethora of 2D structures, however, only very few have been
found to exhibit intrinsic ferroelectric behavior. With few 2D
systems available that show ferroelectric behavior, which moreover
is not very interesting, hope for useful applications of 2D
ferroelectrics started to dwindle. A useful change appeared when
intrinsically non-ferroelectric 2D materials  have been decorated
by functional radicals that induce ferroelectric behavior. This
was first demonstrated when graphene turned ferroelectric
following the functionalization by $OH-$ groups~\cite{wu2013}.
Functionalization by different radicals offers almost unlimited
possibilities to fabricate 2D ferroelectrics. In this study, we
focus on 2D electrides consisting of carbides of lanthanide
elements that are terminated by oxygen.

Electrides form a family of materials, where interstitial anionic
electrons are trapped in cavities within the
sublattice of positive ions~\cite{{Hosono13},{Hosono16},%
{Hosono17},{Hosono18},{Burton20}}. Back in 2013, layered electride
Ca$_2$N was successfully synthesized~\cite{Hosono13} and Ca$_2$N
monolayers were subsequently isolated in 2016 using liquid
exfoliation~\cite{Scott16}. Since then, various 2D layered
electrides have been synthesized experimentally or
predicted theoretically~\cite{{Hosono14},{Sung18},%
{Sung20},{mcrae22},{huang181},{park171},{horiba17},{qiu22},{gui23}}.
However, these layered electrides are only stable under very
specific conditions. These include a nitrogen atmosphere or
selected organic solvents for monolayers of
Ca$_2$N~\cite{Scott16}, ultrahigh vacuum conditions for
Y$_2$C~\cite{Sung18}, and high-purity Ar environment for
Gd$_2$C~\cite{Sung20}. Presence of highly delocalized electrons,
which are trapped in-between individual layers, make these
materials very reactive. Exposure to air leads to undesirable
property changes due to the oxidation of
$-Cl$, $-OH$, $-H$, $-F$ functional groups~\cite{{wang191},%
{zhou181},{hong16},{li161},{hu141},{khazaei16},{maeda20},%
{baghini22},{wang211},{Kai22},{Bu23}}. Sensitivity to specific
environments has also been observed regarding magnetic
polarization. For example, $-Cl$ or $-Br$ termination of Gd$_2$C
causes anti-ferromagnetism with out-of-plane N\'eel ordering,
whereas termination by $-I$ leads to a zigzag anti-ferromagnetic
ordering~\cite{Bu23}. Not only magnetic, but also electronic
properties can be modified and tuned by different functional
groups. Monolayers of Gd$_2$C~\cite{Kai22} become semi-metallic
when covered by hydrogen on one side, but turn insulating when
hydrogen covers both sides. With this richness in behavior, we
expect that ferroelectricity can be induced in many 2D compounds
by specific functional groups.


Our study investigates previously unexplored carbides of
lanthanide elements with the composition M$_2$C (M=Gd, Tb, Dy),
which form 2D structures. Our {\em ab initio} calculations show
that termination of these 2D compounds by oxygen to M$_2$CO$_2$
turns these systems into ferroelectrics. We find two stable phases
of these systems with different ferroelectric properties. The
ferroelectric $\alpha-$phase with out-of-plane polarization is
generally more stable than the anti-ferroelectric $\beta-$phase.
We focus mainly on the $\alpha-$phase, which may be of interest
for applications, and discuss its relation to the $\beta-$phase
with a sublattice of flipped dipoles.
Semiconducting $\alpha-$M$_2$CO$_2$ has an indirect fundamental
band gap that can turn to a direct gap by applying specific
in-plane biaxial strain, which also monotonically changes the
polarization. Our results also reveal excellent photo-absorption
capability of the direct-gap M$_2$CO$_2$ system, which increases
its value for applications.

\section{Computational Techniques}

Our calculations of the atomic structure, electronic and optical
properties have been performed using the Density Functional Theory
(DFT) as implemented in the {\textsc{VASP}}~\cite{VASP,VASPPAW}
code. We used the Perdew-Burke-Ernzerhof (DFT-PBE)~\cite{DFT-PBE}
exchange-correlation functional for most of the study. Specific
calculations of the band structure and the optical absorption
coefficient were performed using the hybrid DFT-HSE06
functional~\cite{{Ernzerhof03},{Ernzerhof06}} with the default
mixing parameter $\alpha = 0.25$. Periodic boundary conditions
have been used throughout the study, with monolayers represented
by a periodic array of slabs separated by a 20~{\AA} thick vacuum
region. The calculations were performed using the projector
augmented wave (PAW) method~\cite{VASPPAW} and a $520$~eV energy
cutoff. The 2D Brillouin zone (BZ) in the reciprocal space has
been sampled by a fine $10{\times}10{\times}1$~$k$-point grid %
\modR{%
in the $\alpha$-phase and $10{\times}6{\times}1$~$k$-point grid in
the $\beta$-phase%
}%
~\cite{Monkhorst-Pack76}. All geometries have been optimized using
the conjugate gradient (CG) method~\cite{CGmethod}, until none of
the residual Hellmann-Feynman forces exceeded $10^{-2}$~eV/{\AA}.
The polarization was calculated using the standard Berry phase
approach~\cite{{Vanderbilt93},{Resta94}} as implemented in the
{\textsc{VASP}} code. The phonon spectrum was determined using
$4{\times}4{\times}1$ supercells, and the real-space force
constants in the supercells were calculated using the
density-functional perturbation theory (DFPT) as implemented in
VASP~\cite{togo2015}.

\begin{figure}[t]
\includegraphics[width=1.0\columnwidth]{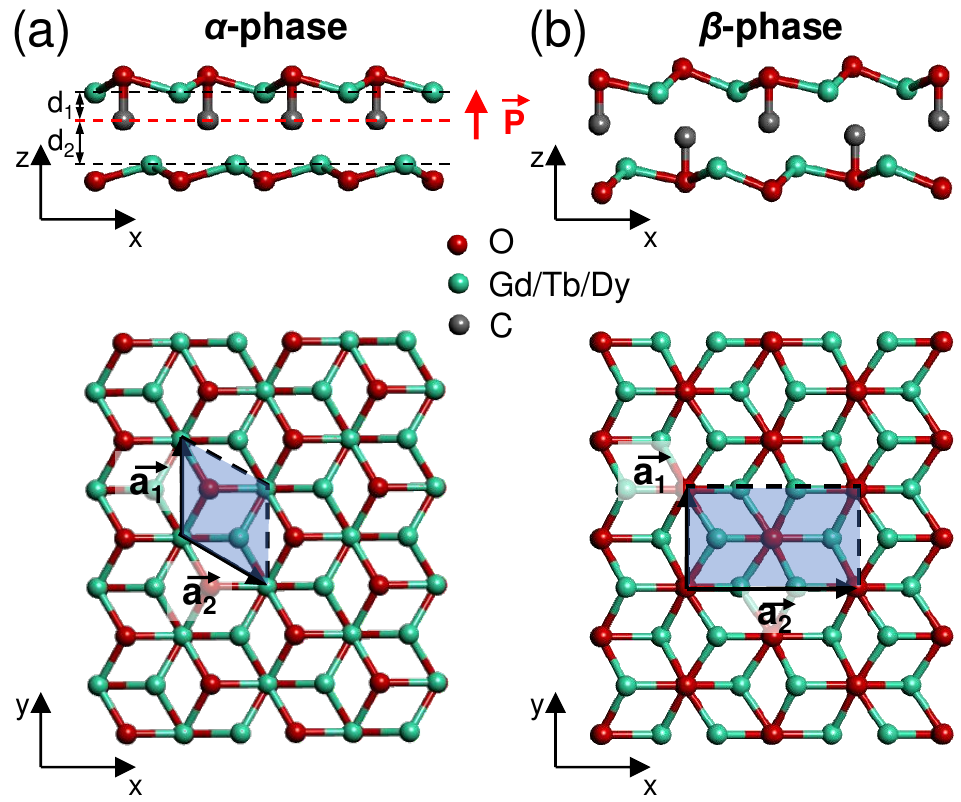}
\caption{%
Atomic structure of 2D M$_2$CO$_2$ (M=Gd, Tb, Dy) in the %
(a) $\alpha-$ and %
(b) $\beta-$phase. %
The unit cells of the 2D structures are highlighted by the
transparent blue areas. $a=a_{1}=a_{2}$ is the length of the
equally long Bravais lattice vectors in the $\alpha-$phase.
The direction of the electric polarization is indicated by the
red arrows. %
\label{fig1}}
\end{figure}

\begin{figure}[t]
\includegraphics[width=1.0\columnwidth]{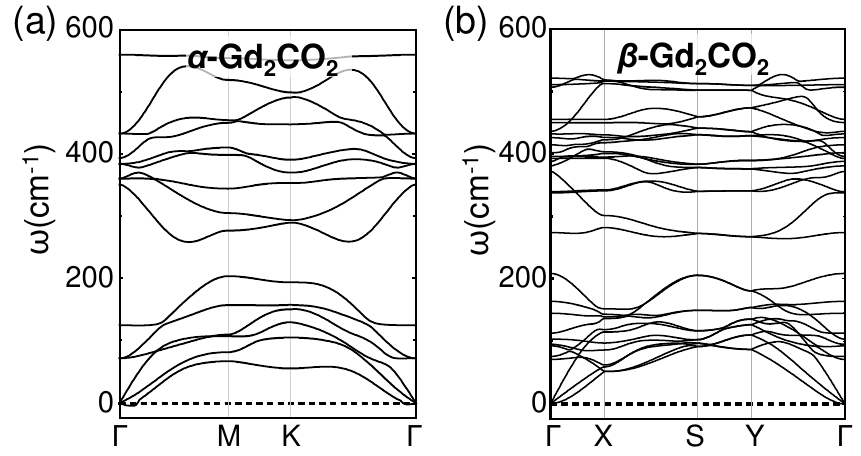}
\caption{%
Phonon spectrum of %
(a) $\alpha-$Gd$_2$CO$_2$ and %
(b) $\beta-$Gd$_2$CO$_2$. %
\label{fig2}}
\end{figure}

\begin{figure*}[t]
\includegraphics[width=1.8\columnwidth]{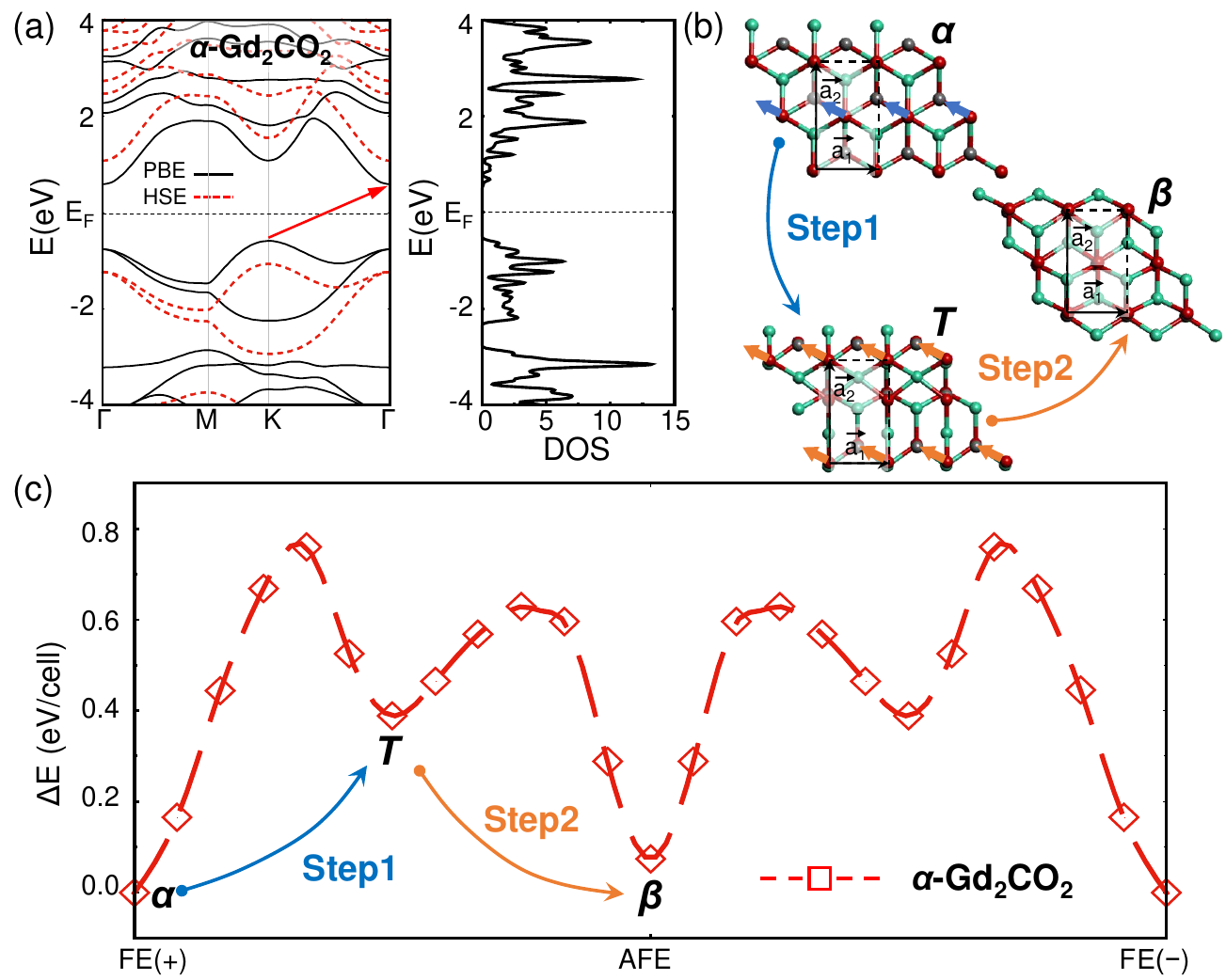}
\caption{%
(a) Electronic band structure $E(k)$ and density of state (DOS) %
of $\alpha-$Gd$_2$CO$_2$. %
Results based on DFT-PBE are represented by the solid black lines and %
DFT-HSE06 results by the dashed red lines. %
(b) Illustration of the transformation process from $\alpha-$Gd$_2$CO$_2$ %
to $\beta-$Gd$_2$CO$_2$. Starting from the initial $\alpha-$phase, displacement %
of O atoms, indicated by the green arrows, leads to the %
transition phase ($T$-phase). Subsequent displacement %
of the O atoms along the direction of the yellow arrows moves the system %
to the $\beta-$phase. %
(c) Energy changes ${\Delta}E$ per unit cell during the transition
process that %
\modR{%
flips dipoles and reverses the electric polarization. %
}%
Ferroelectric FE(+) and FE(-) states are energetically degenerate
eigenstates with opposite polarization direction.%
\label{fig3}}
\end{figure*}

\section{Results}

\subsection{Atomic structure and stability of $O-$terminated
            2D lanthanide carbides M$_2$CO$_2$}

We find $O-$terminated 2D lanthanide carbide structures with the
composition M$_2$CO$_2$ (M=Gd, Tb, Dy) to be stable. Specific
results for Gd-based systems are presented in the main manuscript
and those for Tb and Dy based systems in the Appendix.

Thin slabs of Gd$_2$C contain three atomic layers. Gd atoms form
the top and the bottom layers, and are separated equidistantly by
a C layer in the middle. Since the Gd$_2$C layer is chemically
unstable in air~\cite{Sung20}, we studied a potentially more
stable compound formed by oxidizing both surfaces of Gd$_2$C. We
found two stable allotropes of Gd$_2$CO$_2$, namely the
$\alpha-$phase shown in Fig.~\ref{fig1}(a) and the $\beta-$phase
shown in Fig.~\ref{fig1}(b). Since Tb and Dy are neighbors of Gd
in the lanthanoid series, their compounds show similar chemical
behavior and physical properties as those of Gd.

The optimum lattice constant in the $\alpha-$phase is
$a=a_{1}=a_{2}=3.77$~{\AA}. In the $\beta-$phase, the optimum
lattice constants are $a_{1}=3.70$~{\AA} and $a_{2}=6.44$~{\AA}.
Based on DFT-PBE results, the $\alpha-$phase of M$_2$CO$_2$ is
more stable than the $\beta-$phase by 29~meV/unit-cell for M=Gd,
26~meV/unit-cell for M=Tb, and 25~meV/unit-cell for the Dy-based
compound. In all systems, the C atoms have moved from their
initial central position in-between the M layers. In the
$\alpha-$phase, all of the C atoms bond to O atoms in one of the
terminating layers and acquire a net positive charge. The carbon
layer remains planar, but is no longer equidistant to the
terminating M layers. As seen in the top panel of
Fig.~\ref{fig1}(a), its separation $d_{1}$ from one terminating M
layer is shorter than the separation $d_{2}$ from the other
terminating layer. We characterize the deviation from equidistance
by the quantity $\Delta{d}=(d_{2}-d_{1})/2$. The $\alpha-$phase
shows a $P3m1$ symmetry and exhibits a net out-of-plane electric
polarization.

In the $\beta-$phase with $P2_1/m$ symmetry, the C atoms in the
middle layer bond alternatively to oxygen atoms in one or the
other terminating M layer. As seen in the upper panel of
Fig.~\ref{fig1}(b), the carbon layer is no longer planar. The
local polarization changes along the alternating displacement
direction of the C atoms. As a result, the $\beta-$phase exhibits
zero net polarization.

We also calculated the phonon spectra of the two Gd$_2$CO$_2$
phases and present the phonon band structure in Fig.~\ref{fig2}.
The frequency spectrum of both phases is free of imaginary
frequencies, which means that both allotropes are dynamically
stable. We reach the same conclusion about the dynamical stability
of Tb$_2$CO$_2$ and Dy$_2$CO$_2$ based on their phonon spectra
that are presented in Appendix A.

In the following text, we will focus on the electronic and the
optical properties of %
\modR{%
M$_2$CO$_2$ structures %
}%
in the $\alpha-$phase, which is more stable and exhibits non-zero
electric polarization. %
\modR{%
As we discuss in Appendices A and B, we find $\alpha$-M$_2$CO$_2$
systems to be not only thermodynamically stable at room
temperature, but also dynamically stable even under moderate
compressive strain. %
}

\begin{figure}[ht]
\includegraphics[width=0.8\columnwidth]{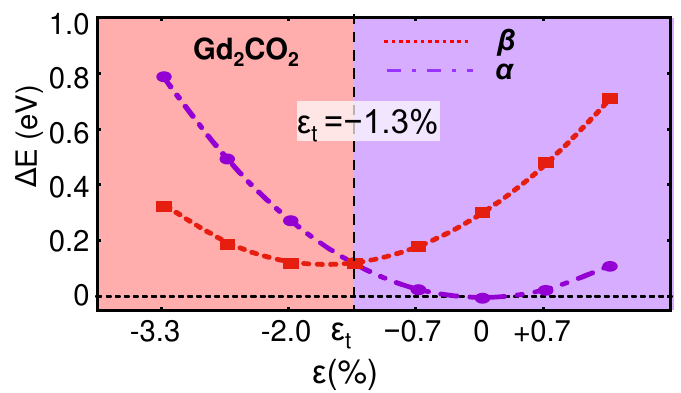}
\caption{%
Energy change $\Delta{E}$ as function of in-plane biaxial strain
$\varepsilon$ applied to the $\alpha-$ and $\beta-$phases of
Gd$_2$CO$_2$, with $\varepsilon=0$ referring to
$\alpha-$Gd$_2$CO$_2$ in equilibrium. We emphasize the region,
where the $\alpha-$phase is more stable, by the purple background
and the region, where the $\beta-$phase is more stable, by the red
background. The dashed line at $\varepsilon_{t}$ separates the
regions, where both phases
are equally stable. %
\label{fig4}}
\end{figure}

\begin{figure}[b]
\includegraphics[width=1.0\columnwidth]{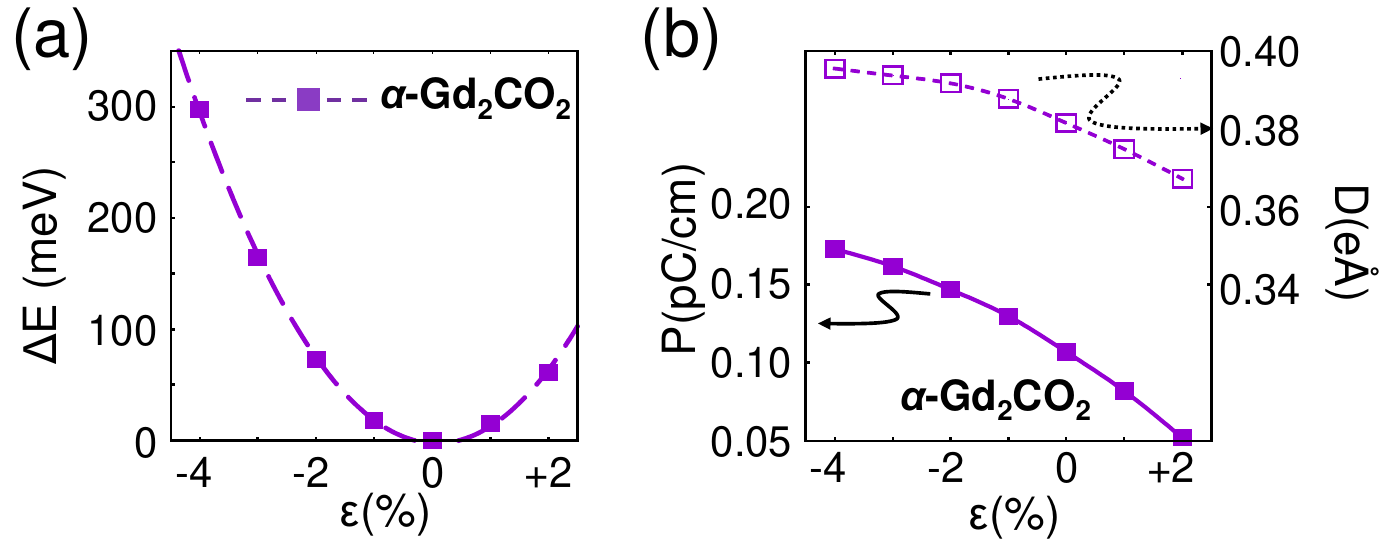}
\caption{%
(a) Energy changes $\Delta{E}$ as function of in-plane biaxial %
strain $\varepsilon$ applied to $\alpha-$Gd$_2$CO$_2$. %
(b) Changes in the electric polarization $P$ and the electric %
dipole $D$ as function of in-plane biaxial %
strain $\varepsilon$ applied to $\alpha-$Gd$_2$CO$_2$. %
\label{fig5}}
\end{figure}

\subsection{Electronic structure and polarization behavior of
            $\alpha-$Gd$_2$CO$_2$}

Metallic behavior of the layered 2D electride Gd$_2$C originates
from itinerant electrons accumulated in-between layers.
Functionalizing the terminating layers with O engages unpaired
electrons in new chemical bonds with the central C atoms. This
changes the electronic structure of the system that is now free of
itinerant electrons. As shown in Fig.~\ref{fig3}(a), DFT
calculations indicate that $\alpha-$Gd$_2$CO$_2$ turns to a
semiconductor. We find the valence band maximum (VBM) at the
$K$-point and the conduction band minimum (CBM) at the
$\Gamma$-point, indicating an indirect band gap.

We should note here that the interpretation of Kohn-Sham
eigenvalues as self-energies is strictly incorrect and that
DFT-based band gaps are typically underestimated. We find this to
be the case also when comparing band gap values based on DFT-PBE
and the hybrid DFT-HSE06 functional, which is believed to be
superior to DFT-PBE in terms of optical spectra. The band gap of
$\alpha-$Gd$_2$CO$_2$ is 1.17~eV based on DFT-PBE and 2.15~eV
based on DFT-HSE06. As seen in Fig.~\ref{fig3}(a) and as expected,
the band dispersion is the same in both functionals. We find the
same to be true also for the band structure of
$\alpha-$Tb$_2$CO$_2$ and $\alpha-$Dy$_2$CO$_2$ that are discussed
in Appendix C.

As mentioned above, the positively charged C atoms are displaced
from their center position towards one of the terminating Gd
layers in $\alpha-$Gd$_2$CO$_2$. As a result, the system acquires
a macroscopic electric polarization in the out-of-plane direction.
We use the standard Berry-phase method to calculate the
polarization and obtain $P$=0.107~pC/cm for $\alpha-$Gd$_2$CO$_2$.
Similarly, we get $P$=0.113~pC/cm for $\alpha-$Tb$_2$CO$_2$ and
$P$=0.148~pC/cm for $\alpha-$Dy$_2$CO$_2$.

\subsection{Energy barrier for polarization reversal in
            ferroelectric $\alpha-$Gd$_2$CO$_2$}

Since the origin of polarization in the $\alpha-$phase is a
uniform displacement of the central C layer towards one of the
terminating layers, the polarization can be reversed when the C
layer is displaced uniformly towards the other terminating layer,
as can be simply inferred from Fig.~\ref{fig1}(a). The possibility
to reverse the polarization makes the $\alpha-$phase a
ferroelectric. Displacing alternatively every other C atom in one
or the other direction leads to the $\beta-$phase, shown in
Fig.~\ref{fig1}(b). The absence of net polarization in spite of
nonzero alternating dipole moments makes the $\beta-$phase an
anti-ferroelectric.

As mentioned, the polarization of the $\alpha-$phase could be
reversed by displacing all C atoms at once, yielding an
energetically degenerate $\alpha-$phase with opposite
polarization. Rather than displacing all C atoms at once, the
carbons could be shifted up or down alternatively, resulting in
the $\beta-$phase as a locally stable transition state. We
illustrate this process from the $\alpha-$ to the $\beta-$phase in
Fig.~\ref{fig3}(b) using the unit cells that contain four C and
four O atoms. Starting with the $\alpha-$phase, this transition
starts by displacing only one O towards its closest C neighbor and
attracting it to form a metastable state $T$. Similar displacement
of the second O atom in the unit cell causes a transition to the
$\beta-$phase. The energy barrier ${\Delta}E$ associated with the
polarization reversal process in $\alpha-$Gd$_2$CO$_2$ has been
determined using the nudged elastic band (NEB) method and the
results are shown in Fig.~\ref{fig3}(c). %

\modR{%
The barrier value $757$~meV/unit cell for polarization reversal in
Gd$_2$CO$_2$, which has been obtained in this way, is rather high.
Even though this value is per unit cell and not per atom, it is
significantly larger than $k_BT{\approx}0.02$~eV/atom and thus
should protect the polarization direction at room temperature.
Applying an external field beyond the coercive field, a common way
to reverse electric polarization in dielectrics, is unlikely to
succeed in view of this high energy barrier. Not to dismiss this
system as a dielectric based on the energy barrier alone, we wish
to refer to a different system, a LiNbO$_3$-type corundum
derivative FeTiO, with an even higher activation barrier of
$763$~meV/unit cell. In spite of this high barrier, ferroelectric
behavior has been postulated in a theoretical study~\cite{DV16}
and later on confirmed experimentally~\cite{PRL09}. %
}

\modR{%
The approach we used to determine the activation barrier for
polarization reversal is based on an artificial coherent process
extending to the entire system, where every unit cell undergoes
the change simultaneously. This view is oversimplified and clearly
overestimates the barrier height to values unsurmountable at room
temperature. As we expand in the Discussion Section, the barrier
in a realistic system, where the transition proceeds by defect
motion that modifies domains, is significantly lower. %
}

\modR{%
In Gd$_2$CO$_2$ and other related systems we consider here, we
found that applied strain may change the energetically favored
phase. As shown in Fig.~\ref{fig4}, the energetically favored
phase of Gd$_2$CO$_2$ changes from $\alpha$ to $\beta$ under
compressive strain $\varepsilon<\varepsilon_{t}$. Since the
transition strain value $\varepsilon_{t}{\approx}-1.3$\% in
Gd$_2$CO$_2$ is relatively moderate, the possibility of phase
stability reversal needs careful consideration when determining
the effect of strain on the band structure, optical absorption and
ferroelectric behavior of the system. }

\modR{%
Even though the ferroelectric $\alpha-$phase becomes less stable
than the $\beta-$phase in the compression range associated with a
direct band gap, it is protected from a spontaneous transition to
the $\beta-$phase by the above-mentioned high activation barrier.
We have confirmed that even under compression, this activation
barrier remains high, so that the electronic and optical
properties of ferroelectric $\alpha-$Gd$_2$CO$_2$ are can be
depended upon. %
}

\modR{%
We have performed analogous calculations also for the other
M$_2$CO$_2$ systems, Tb$_2$CO$_2$ and Dy$_2$CO$_2$. We report
polarization changes in these systems in Appendix C and stability
differences between the $\alpha$ and the $\beta$ phase in Appendix
D. We find these results to be very similar to those for
Gd$_2$CO$_2$. }

\subsection{Effect of in-plane strain on the electronic properties and
            electric polarization of $\alpha-$Gd$_2$CO$_2$}

It is well known that lattice distortion can modify physical
properties of materials. Applying external strain is a common way
to modify the lattice structure. We use the common definition of
strain $\varepsilon={\Delta}a/a$. In this study, we apply in-plane
biaxial strain on the 2D Gd$_2$CO$_2$ monolayer, with
$-4\%<\varepsilon<+2\%$, in order to modify the the electronic
band structure and the electric polarization. Before distorting
the lattice, we investigated the energy needed to apply the strain
and present our results in Fig.~\ref{fig5}(a). Applying
compressive strain of $\varepsilon=-4\%$ requires
$\Delta{E}{\approx}300$~meV/unit~cell and applying tensile strain
of $\varepsilon=+2\%$ cost $\Delta{E}{\approx}60$~meV/unit~cell.
These values are close to those found in layered bulk black
phosphorus~\cite{DT254}, indicating similar mechanical properties.

\begin{figure}[t]
\includegraphics[width=1.0\columnwidth]{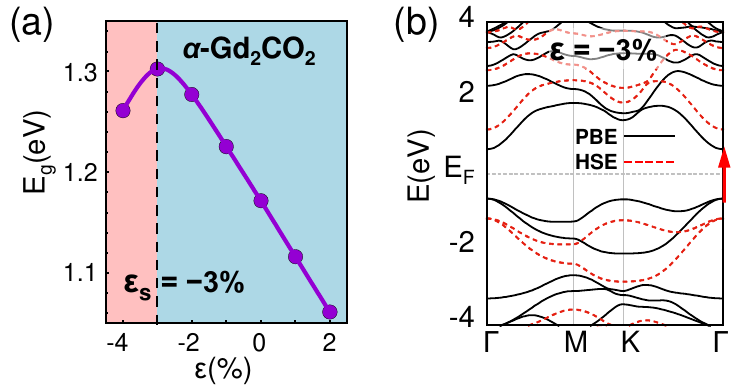}
\caption{%
(a) Dependence of the electronic band gap $E_{g}$ on the in-plane
biaxial strain $\varepsilon$. Compression beyond a specific value
$\varepsilon_{s}<0$ turns the system to a direct gap
semiconductor. The direct gap semiconducting region
$\varepsilon<\varepsilon_{s}$ is highlighted by pink and the
indirect band gap region by light
blue. %
(b) The electronic band structure $E(k)$ of $\alpha-$Gd$_2$CO$_2$
strained at $\varepsilon=\varepsilon_{s}$. Results based on the
DFT-PBE functional are represented by the solid black line and
results based on the DFT-HSE06 functional by the dashed red line. %
\label{fig6}}
\end{figure}

\begin{figure}[t]
\includegraphics[width=1.0\columnwidth]{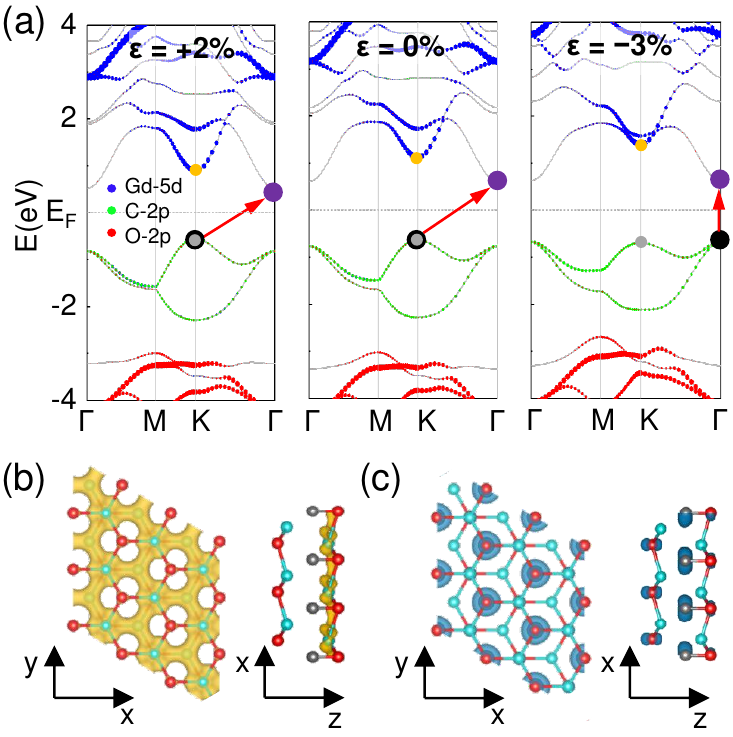}
\caption{%
(a) \modR{Orbital-projected} electronic band structure $E(k)$ of
$\alpha-$Gd$_2$CO$_2$ stretched by %
$\varepsilon=+2\%$,
relaxed at %
$\varepsilon=0\%$, and compressed by $\varepsilon=-3\%$. The
CBM is indicated by the %
\modR{%
purple %
}%
point and the VBM by the %
\modR{%
black %
}%
point. At the $K$-point, the conduction band eigenvalue is
indicated by the yellow point and the valence band eigenvalue
by the %
\modR{%
grey %
}%
point. %
Isosurface charge density plots at the value
$\rho=0.035~e/a_B^{3}$ represent the distribution of %
(b) conduction band electrons, shown in yellow, and %
(c) valence band electrons, shown in blue, %
at the $K$-point in an unstrained structure. %
\label{fig7}}
\end{figure}

Depending on applied strain, we find the off-center displacement
$\Delta{d}$ of C atoms in the middle layer to change along with
the net Bader charge $Q$(C) of the C atoms. Electrons transferred
from C to O atoms form an electrical dipole. We define this dipole
by $D=Q$(C)${\times}\Delta{d}$ and present its dependence on the
strain in Fig.~\ref{fig5}(b). Independent of the estimated dipole
value, we calculate the polarization $P$ using the Berry phase
method. We note different units, but a similar dependence of $P$
and $D$ on the strain. There may be additional difference between
the two quantities, since both the Bader charge analysis for $D$
and the Berry phase method for $P$ are approximations. As seen in
Fig.~\ref{fig5}(b), changing the strain from the tensile value
$\varepsilon=+2\%$ to compression at $\varepsilon=-4\%$ causes a
$333\%$ increase of the polarization.

Also the electronic band structure is susceptible to biaxial
in-plane strain. As seen in Fig.~\ref{fig6}(a), the band gap
increases under compressive strain and decreases under tensile
strain. At the specific compressive strain value
$\varepsilon=\varepsilon_{s}=-3$\%, the band gap reaches its
maximum and $\alpha-$Gd$_2$CO$_2$ turns from an indirect-gap to a
direct-gap semiconductor. The band structure of
$\alpha-$Gd$_2$CO$_2$ subject to $\varepsilon_{s}$ strain is shown
in Fig.~\ref{fig6}(b). In comparison to the band structure of the
unstrained system in Fig.~\ref{fig3}(a), the CBM remains at the
$\Gamma$-point, but the VBM moves from the $K$-point to the
$\Gamma$-point under in-plane compression. We observe the gap to
decrease upon compression beyond $\varepsilon_{s}$, the switching
point to a direct-gap semiconductor. The effect of biaxial strain
on the band structure and polarization of $\alpha-$Tb$_2$CO$_2$
and $\alpha-$Dy$_2$CO$_2$ is discussed in Appendix E.

In the following we analyze the mechanism underlying the
transition from an indirect to a direct band gap due to in-plane
strain in $\alpha-$Gd$_2$CO$_2$. We base our analysis on comparing
the %
\modR{%
orbital-projected band structure of $\alpha-$Gd$_2$CO$_2$ at
strain values $\varepsilon=+2\%$, $\varepsilon=0\%$ and
$\varepsilon=-3\%$ in Fig.~\ref{fig7}(a). In comparison to the
moderate changes at the VBM and CBM at the $\Gamma$-point by
strain, the corresponding changes at the $K$-point are dramatic
and worth further consideration. %
}

\modR{%
As seen in Fig.~\ref{fig7}(a), the character of the VBM and CBM at
the $K$-point consists mostly of C$2p$ and Gd$5d$ states. %
For the unstrained system, we display the $K$-point electron
distribution at the CBM in Fig.~\ref{fig7}(b) and at the VBM in
Fig.~\ref{fig7}(c). We find that CBM states at the $K$-point
consist mostly of hybridized Gd$5d_{xy}$ and Gd$5d_{x^2-y^2}$
orbitals, which hybridize in covalent bonds between neighboring Gd
atoms. The VBM at the $K$-point, on the other hand, consists
mainly of isolated C$2p_{x}$ and C$2p_{y}$ orbitals, which are
responsible for forming covalent bonds between neighboring C
atoms. As we show below, the Gd-Gd bonds in $\alpha-$Gd$_2$CO$_2$
are compressed and C-C bonds stretched with respect to their
equilibrium values, which affects their behavior under strain.
}%

\modR{%
The calculated Gd-Gd interatomic distance $3.77$~{\AA} in
$\alpha-$Gd$_2$CO$_2$ turns out to be much smaller than the
optimum value $d_{\rm{opt}}$(Gd-Gd)$=4.56$~{\AA} found in bulk
hcp-Gd~\cite{zeng2011}, indicating that Gd-Gd bonds in the
oxygenated carbide are under compression. Further compression
should then move these states up in energy, away from the Fermi
level. Still, the CBM remains at $\Gamma$. %
}%

\modR{%
On the other hand, the C-C interatomic distance
$d$(C-C)$=3.77$~{\AA} in Gd$_2$CO$_2$ is much larger than the
equilibrium length of C-C single bonds,
$d_{\rm{opt}}$(C-C)$=1.54$~{\AA}, and C=C double bonds,
$d_{\rm{opt}}$(C=C)$=1.45$~{\AA}~\cite{Beach1935}. Thus, carbon
orbitals barely overlap in the oxygenated carbide and C-C bonds
are under tension. Compression should lower the energy of the VBM,
away from the Fermi level, even below its eigenvalue at $\Gamma$.
At some minimum compressive strain value $\varepsilon_{s}$, the
maximum of the valence band moves from $K$ to $\Gamma$ and the
band gap becomes direct. %
}%

\subsection{Optical properties of $\alpha-$Gd$_2$CO$_2$}

As discussed above, the $\alpha-$Gd$_2$CO$_2$ system can be turned
into a direct-gap semiconductor by applying biaxial in-layer
compressive strain. A direct band gap is highly desirable for
optical applications, as it allows for a direct recombination of
electrons in the conduction band with holes in the valence band,
resulting in a high luminous efficiency. Even though the $E(k)$
dispersion relation is reproduced rather correctly by DFT-PBE, DFT
is not designed to reproduce the fundamental band gap $E_{g}$ and
underestimates it significantly. Thus, for optical properties, we
rely on the more appropriate DFT-HSE functional. Our study of the
photon absorption properties of $\alpha-$Gd$_2$CO$_2$ is based on
electronic structure calculations using the DFT-HSE06 functional.

\begin{figure}[b]
\includegraphics[width=1.0\columnwidth]{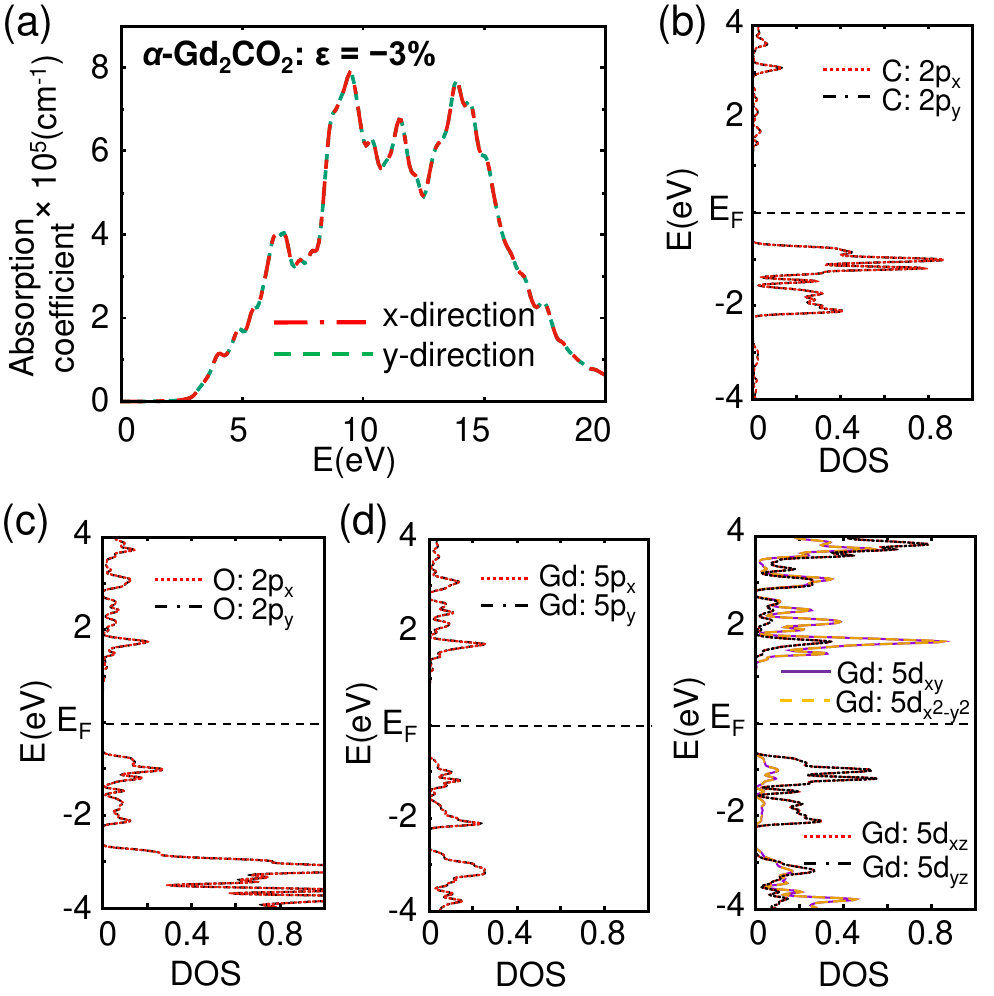}
\caption{%
(a) The optical absorption coefficient of $\alpha-$Gd$_2$CO$_2$
subject to isotropic strain $\varepsilon=-3\%$. %
Electronic density of states projected on
(b) 2$p_{x}$ and 2$p_{y}$ states of C atoms, %
(c) 2$p_{x}$ and 2$p_{y}$ states of O atoms, and %
(d) 5$p_{x}$, 5$p_{y}$, 5$d_{xy}$, %
    5$d_{x^2-y^2}$, 5$d_{xz}$ and 5$d_{yz}$ states of Gd atoms. %
\label{fig8}}
\end{figure}

\modR{%
We determine the optical absorption coefficient $\alpha(\omega)$
from
\begin{equation}
\alpha(\omega)=\sqrt{2}\omega%
\surd[\sqrt{\varepsilon^{2}_{1}+\varepsilon^{2}_{2}} %
-\varepsilon_{1}]/c \;, %
\label{Eq1}
\end{equation}
where $\omega$ is the angular frequency and $c$ is the speed of
light. $\varepsilon_{1}$ is the real and $\varepsilon_{2}$ the
imaginary part of the dielectric function, which can be calculated
using}
\modR{
\begin{eqnarray}
& & \varepsilon_{2}(\omega) = \frac{8{\pi}^{2}e^2}{{\omega}^{2}m^{2}} \times  \\
& & \sum_{n}\sum_{n'} \int {\lvert {P_{nn'}^{k}}\rvert}^2 f_{kn}(1-f_{nn'}) %
\delta(E_{n}^{k}-E_{n'}^{k}-\frac{h{\omega}}{2\pi})
\frac{d^3k}{(2\pi)^3} \nonumber %
\label{Eq2}
\end{eqnarray}
and %
}
\modR{
\begin{equation}
\varepsilon_{1}(\omega) = 1+\frac{2P}{\pi} \int_{0}^{\infty} %
\frac{{\omega}'\varepsilon_{2}(\omega')}{{\omega'}^2-{\omega}^2} d{\omega'} \,.%
\label{Eq3}
\end{equation}
}
\modR{%
Here, $P_{nn'}^{k}$ is the projection of elements of the momentum
dipole matrix, $f_{kn}$ is the Fermi-Dirac distribution and
$E_{n}^{k}(k)$ is the energy of the electron. $m$ is the mass and
$e$ the charge of the electron.%
}

Our results in Fig.~\ref{fig8}(a) indicate that the absorption
threshold of $\alpha-$Gd$_2$CO$_2$ under strain $\varepsilon$
should occur near $2.5$~eV, which is very close to the calculated
band gap. We should expect no optical absorption below the
corresponding frequency. We can identify five strong absorption
peaks in the range from $5-20$~eV. Two of these can be associated
with the ultraviolet spectrum of carbon in the energy range from
$4.4-12.4$~eV. The strongest three peaks occur in the
extreme-ultraviolet range of the optical spectrum ranging from
$10.25-124$~eV.

As seen in Fig.~\ref{fig8}(a), the calculated absorption spectra
$\alpha(\omega)$ are the same in the $x$ and $y$ direction,
indicating that the absorption should be isotropic. %
\modR{%
This is caused by the energetic degeneracy of C$2p_{x}$ and
C$2p_{y}$, Gd$5d_{xy}$ and Gd$5d_{x^2-y^2}$, Gd$5d_{xz}$ and
Gd$5d_{yz}$ states that can be understood using group theory. As
mentioned above, the space group of $\alpha-$M$_2$CO$_2$
structures is $P3m1$ and the point group is $C_{3v}$. The linear
function of ($x$, $y$), quadratic function of ($x^2-y^2$, $xy$)
and ($xz$, $yz$) belong to the $E$ irreducible representation of
the $C_{3v}$ group, which means that $p_{x}$ and $p_{y}$ are
degenerate in energy, same as $d_{xy}$, $d_{x^2-y^2}$, $d_{xz}$
and $d_{yz}$ are degenerate in energy. %
}%

To confirm our interpretation, we analyzed the total density of
states by plotting the partial density of states (PDOS) associated
with the elements present in the structure and display our results
in Figs.~\ref{fig8}(b)-\ref{fig8}(d). Our PDOS results show that
the $p_{x}$ and $p_{y}$ states of C, O and Gd atoms are identical.
Similarly, we find the Gd$5d_{xy}$ and Gd$5d_{x^2-y^2}$, the
Gd$5d_{xz}$ and Gd$5d_{yz}$ eigenstates to be identical. Thus, the
electronic behavior of Gd$_2$CO$_2$ is the same in the $x$ and $y$
direction, meaning it is isotropic.

\modR{%
As further discussed in Appendix F, we also find the absorption
spectra of Tb$_2$CO$_2$ and Dy$_2$CO$_2$ under compression to be
the same in the $x$ and $y$ direction and thus isotropic. To make
sure that these structures do not collapse under the compressive
strain, we first calculated their phonon spectra. As we further
expand in Appendix A, we confirmed the stability of these
structures under compression by absence of imaginary frequencies
in their phonon spectra. %
}

\section{Discussion}

\modR{%
As we know, Gd is a rare-earth element in the lanthanide series.
Since rare earth elements are heavy, we expect spin-orbit coupling
(SOC) to play a significant role in the band structure and
possibly to also affect significantly the optical properties. We
have investigated this point in Appendix G and find that SOC does
split bands, yet the split is very small at the VBM and CBM, which
dominate the optical response. We conclude that SOC effects play
only a minor role in optical properties of the M$_2$CO$_2$
systems of interest here. %
}%

Bulk Gd is ferromagnetic (FM) below the Curie temperature
$T_C=20^{\circ}$C. Also the 2D electride Gd$_2$C turns out to be
ferromagnetic~\cite{Sung20}. Even though we have not touched upon
magnetism in this study of the electronic properties, we still may
ask, whether there may be magnetic order in Gd$_2$CO$_2$. A
related question may be, whether any magnetic order in the system
may affect ferroelectricity. Not providing sufficient depth in a
corresponding study, we considered the possibility of
ferromagnetic (FM) and anti-ferromagnetic (AFM) ordering in the
system.

Assuming a FM order, we found that Gd in Gd$_2$CO$_2$ carries a
magnetic moment of 7~$\mu_B$, which is close to its measured local
moment of $7.26~\mu_B$ in Gd$_2$C~\cite{Gschneidner11}. Assuming
an AFM order, the magnetic moments of Gd atoms in the terminating
layers point in opposite direction. Assuming any of these magnetic
orders, we find the $\alpha-$phase to be still most stable and the
atomic structure to be nearly identical to the nonmagnetic
structure. The magnetic properties of Gd$_2$CO$_2$ originate in
the Gd$4f$ electrons. In comparison the Gd$5d$ and Gd$6s$
electrons, Gd$4f$ electrons reside in the inner shell of the atom
and thus do not contribute to chemical bonding. Thus, the effect
of magnetic order on the atomic structure and also the
ferroelectric behavior can be ignored. Also, the Gd$4f$ electrons
are deep below the Fermi level. Hence, the optical properties are
not affected by the magnetic order to a significant level.

\modR{%
The microscopic transition from the $\alpha-$ to the
$\beta-$phase, depicted in Figs.~\ref{fig1} and \ref{fig3}(b),
involves C atoms in the middle layer. In the Gd$_2$C system, these
atoms are equidistant to Gd atoms in the terminating layers. This
situation, however, changes with oxidation. %
}%
As O atoms attach to the Gd atoms, the single-well potential of
the C atoms changes to a double-well potential and the system
gains energy, when the C atoms move closer to one or the other
terminating layer. This symmetry breaking results in the formation
of electric dipoles that are responsible for ferroelectricity and
change the system to an indirect gap semiconductor.

We expect that the transition from a single-well to a double-well
potential is not limited to oxygen termination, but may well occur
with other functional groups such as hydroxyls. Unlike oxygen
atoms, hydroxyl groups carry an electric dipole that may, upon
proper alignment, give rise to in-plane ferroelectricity. While
this consideration exceeds the scope of our study, we anticipate
that combining M$_2$C with other functional groups may lead to
other types of behavior of interest for applications.

\modR{%
Next we wish to revisit the high value of the activation barrier
for polarization reversal in the M$_2$CO$_2$ systems. As in all
phase changes, polarization reversal is a complex process that
usually involves nucleation and motion of defects and requires
significantly lower activation barriers than those estimated for
the artificial coherent process considered here. Whereas the
nature of specific defects is a complex problem beyond the scope
of our study, this topic has been discussed before in a different,
simpler system and context. In elemental selenium, so-called
`point-dislocation motion' has been shown to significantly lower
the activation barrier for the transformation of a one-dimensional
Se helix to a two-dimensional allotrope~\cite{DT267}.%
}%

\modR{%
Similar to the example of defective Se, the M$_2$CO$_2$ systems
have been formed at nonzero temperature. They will also be
non-uniform, but rather contain of coexisting $\alpha-$ and
$\beta-$ phase domains, separated by domain walls. At whatever
applied strain value, the fraction of the dominant phase will
depend on the free energy difference. The transition barrier value
we calculated would matter only in the artificial pure phase. In
the common case of phase coexistence, we expect significant
lowering of the energy barriers for moving dislocations along
domain wall boundaries and will change the fraction of the
dominant phase according to free energy differences. %
}

\modR{%
Unlike in many other electrides, we report in Fig.~\ref{fig4} that
external strain may change the energetically favored phase from
the ferroelectric $\alpha-$M$_2$CO$_2$ to the antiferroelectric
$\beta$ phase. In our opinion, this interesting aspect of the the
system may be used in our favor. Even though each phase is
protected by a substantial activation barrier, we expect this
barrier to generally decrease in a strained system. We can imagine
that subjecting the $\alpha-$phase temporarily to a specific
compressive strain may accelerate a transition to the
$\beta-$phase and, eventually, to a ferroelectric $\alpha-$phase
with reversed polarization under applied electric field. Lowering
the ambient temperature and adjusting the applied strain should
then stabilize $\alpha-$M$_2$CO$_2$ with the reversed polarization. %
}

\modR{%
Since our study is a theoretical prediction for M$_2$CO$_2$
structures that have not been synthesized yet, we may only
speculate and provide hopefully valuable hints to experimentalists
interested in procuring them. Except for the successful synthesis
of bulk M$_2$C~\cite{{Kai22},{Tb2C}}, there is only scarce
information available on how these structures may behave in
different environments. We feel that M$_2$C monolayers may behave
similar to structurally and chemically related MXenes, which have
been investigated extensively. Traditional etching with HF has
been shown to lead to $-F$ termination~\cite{Lis18}, and
alkalization to lead to $-OH$ termination~\cite{xieyue14}. Heating
under an inert atmosphere has converted $-OH$ termination to $-O$
termination~\cite{Lu2019}. Thus, we feel that techniques used to
oxidize MXenes may also be used to form M$_2$CO$_2$ structures.%
}

\begin{figure}[b]
\includegraphics[width=1.0\columnwidth]{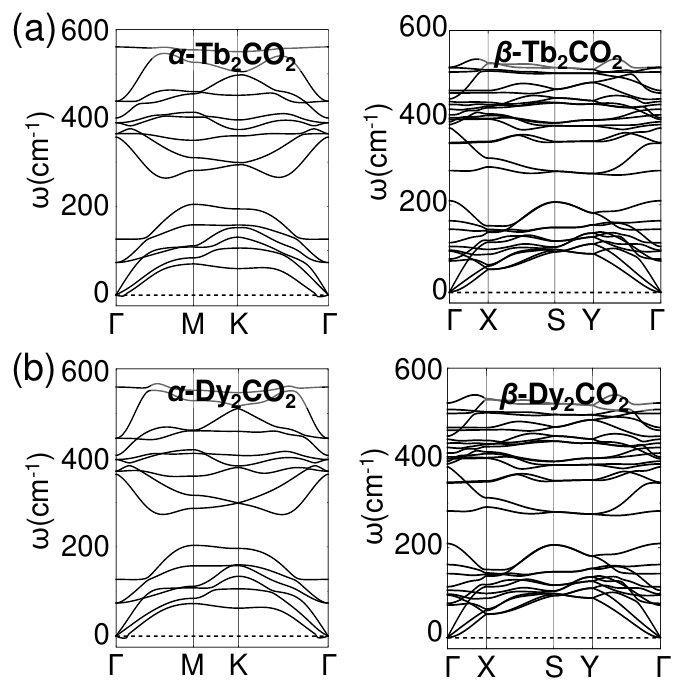}
\caption{%
Phonon spectra of the equilibrated %
$\alpha-$ and $\beta-$phases of %
(a) Tb$_2$CO$_2$, and %
(b) Dy$_2$CO$_2$. %
\label{fig9}}
\end{figure}

\begin{figure}[t]
\includegraphics[width=1.0\columnwidth]{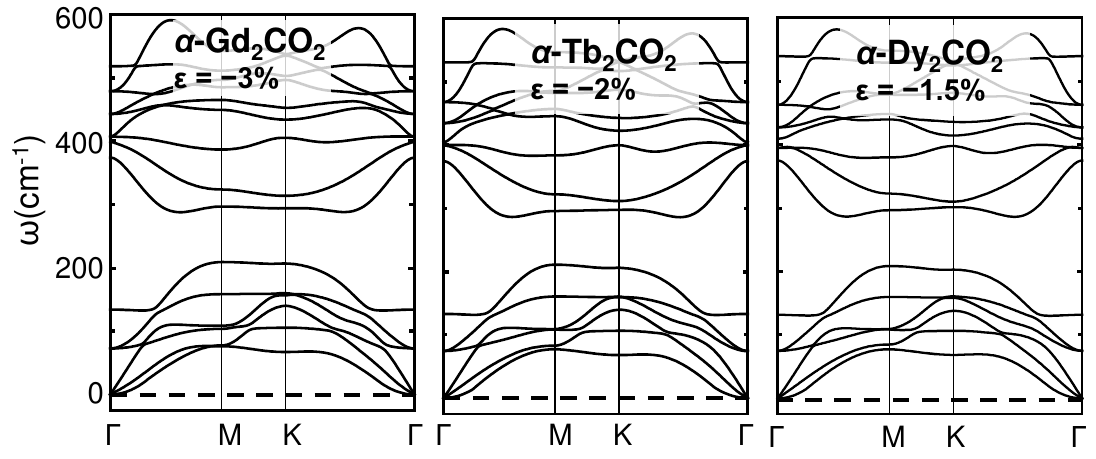}
\caption{%
Phonon spectra of 2D M$_2$CO$_2$ systems under compressive strain
$\varepsilon<0$. Results are presented for %
(a) $\alpha-$Gd$_2$CO$_2$ at $\varepsilon=-3$\%,%
(b) $\alpha-$Tb$_2$CO$_2$ at $\varepsilon=-2$\%, and %
(c) $\alpha-$Dy$_2$CO$_2$ at $\varepsilon=-1.5$\%. %
\label{fig10}}
\end{figure}

\begin{figure}[b]
\includegraphics[width=1.0\columnwidth]{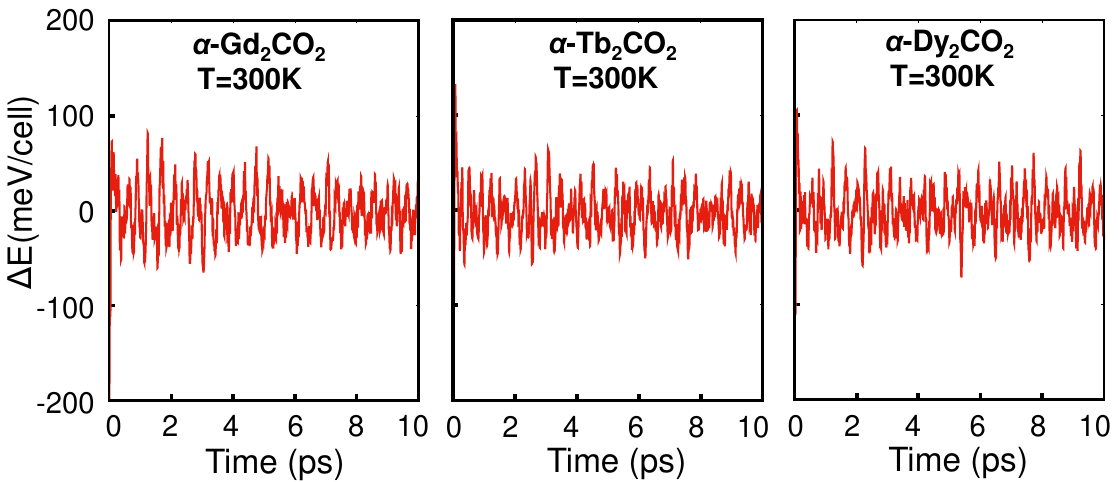}
\caption{%
\modR{%
Changes in the potential energy ${\Delta}E$ per unit cell of %
(a) $\alpha-$Gd$_2$CO$_2$, %
(b) $\alpha-$Tb$_2$CO$_2$, and %
(c) $\alpha-$Dy$_2$CO$_2$ %
during a $10$-ps long canonical MD simulation run at $T=300$~K. %
}%
\label{fig11}}
\end{figure}

\section{Summary and Conclusions}

In this study, we investigate the physical properties of
oxygen-terminated carbides of lanthanide elements with the
composition M$_2$CO$_2$, which form two-dimensional (2D)
structures. Our calculations reveal two dynamically stable phases
of these compounds, namely the energetically favored ferroelectric
$\alpha-$phase with an out-of-plane polarization, and the
anti-ferroelectric $\beta-$phase. Applying in-plane biaxial strain
changes the ferroelectric polarization of the $\alpha-$phase in a
linear fashion, and modifies the size and nature of the
fundamental band gap from direct to indirect. We also find that
the relative stability of the $\alpha-$ and the $\beta-$phase can
be changed by applying in-plane biaxial strain. In structures with
a direct band gap, found in the compressive regime, the in-plane
electronic properties are the same in the $x$ and $y$ direction.
Thus, the optical properties turn isotropic and exhibit excellent
photon absorption in the ultraviolet range. We believe that this
previously unexplored class of systems should find unique
applications among optoelectronic materials.

\section*{Appendix}

\renewcommand\thesubsection{\Alph{subsection}}
\renewcommand{\theequation}{A\arabic{equation}}
\setcounter{subsection}{0} %
\setcounter{equation}{0} %
\setcounter{video}{0} %

\begin{figure}[t]
\includegraphics[width=1.0\columnwidth]{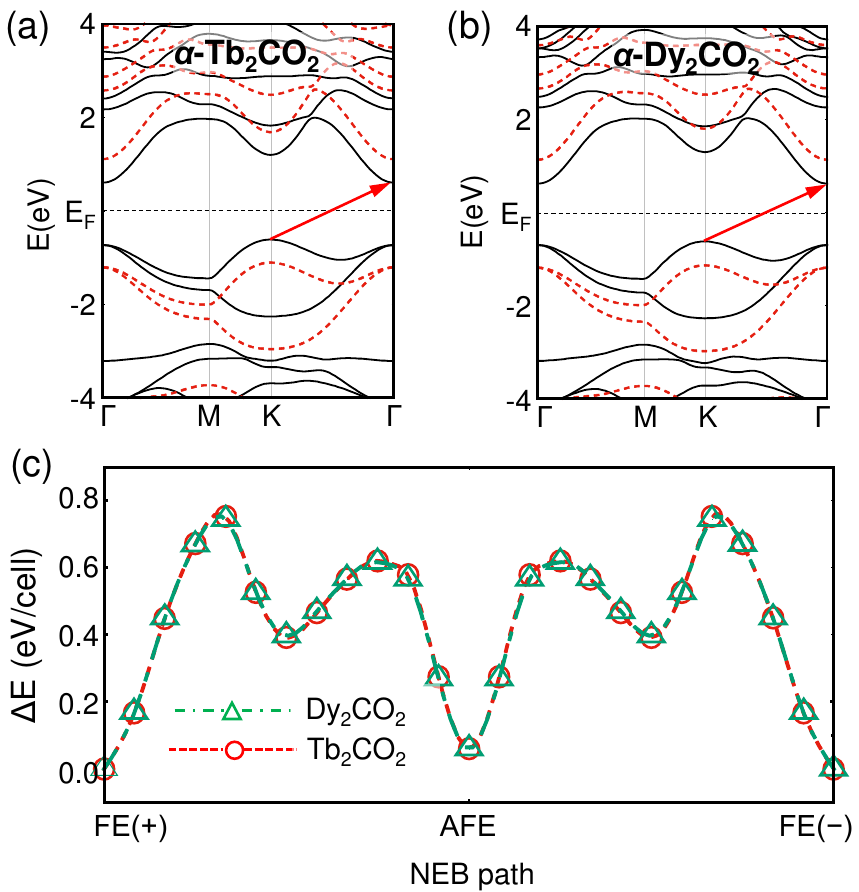}
\caption{%
Electronic band structure $E(k)$ of %
(a) $\alpha-$Tb$_2$CO$_2$ and %
(b) $\alpha-$Dy$_2$CO$_2$. %
DFT-PBE results are represented by the solid black lines and %
DFT-HSE06 results by the dashed red lines. %
(c) Energy changes ${\Delta}E$ per unit cell during the transition
process that flips dipoles and reverses the polarization of
$\alpha-$Tb$_2$CO$_2$ and %
$\alpha-$Dy$_2$CO$_2$. The process is very similar to
$\alpha-$Gd$_2$CO$_2$, as seen in Fig.~\ref{fig3}.%
\label{fig12}}
\end{figure}

\subsection{
Dynamic stability of relaxed and compressed Gd$_2$CO$_2$,
Tb$_2$CO$_2$ and Dy$_2$CO$_2$}

\modR{%
Calculated phonon spectra of unstrained Tb$_2$CO$_2$ and
Dy$_2$CO$_2$ in the $\alpha-$ and $\beta-$phase, with the
structures depicted in Fig.~\ref{fig1}, are presented in
Fig.~\ref{fig9}.%
}

\modR{%
Calculated phonon spectra of Gd$_2$CO$_2$, Tb$_2$CO$_2$ and
Dy$_2$CO$_2$ subject to compressive strain in the value range of
interest for for optical properties, namely
$\varepsilon=-3\%$   in Gd$_2$CO$_2$, %
$\varepsilon=-2\%$   in Tb$_2$CO$_2$, and %
$\varepsilon=-1.5\%$ in Dy$_2$CO$_2$, %
are presented in Fig.~\ref{fig10}.%
}

\modR{%
All spectra are free of imaginary frequencies, meaning that
irrespective of compression, the 2D M$_2$CO$_2$ (M=Gd, Dy and Tb)
structures are dynamically stable.%
}

\subsection{Thermodynamic stability of M$_2$CO$_2$ in the
            $\alpha$-phase}

\modR{%
To study the thermodynamic stability of M$_2$CO$_2$ structures, we
have performed canonical molecular dynamics (MD) simulations of
the $\alpha-$phase at room temperature ($T=300$~K) for a time
period of $10$~ps. As seen in Fig.~\ref{fig11}, the potential
energy fluctuates around a constant value, indicating that all
structures are dynamically stable at room temperature.%
}

\subsection{Band structure and polarization changes in
            Tb$_2$CO$_2$ and Dy$_2$CO$_2$}

Results of our band structure calculations with the DFT-PBE and
DFT-HSE06 functionals are shown in Fig.~\ref{fig12}(a) for
$\alpha-$Tb$_2$CO$_2$ and in Fig.~\ref{fig12}(b) for
$\alpha-$Dy$_2$CO$_2$. Our results indicate that both systems are
indirect-gap semiconductors. The DFT-PBE-based band gap is surely
underestimated at $1.22$~eV for $\alpha-$Tb$_2$CO$_2$ and
$1.24$~eV for $\alpha-$Dy$_2$CO$_2$. Values based on DFT-HSE06,
$2.22$~eV for $\alpha-$Tb$_2$CO$_2$ and $2.27$~eV for
$\alpha-$Dy$_2$CO$_2$, are likely better estimates for the optical
band gap.

The energetics of the process that flips dipoles and reverses the
polarization, presented in Fig.~\ref{fig3}(b) for
$\alpha-$Gd$_2$CO$_2$, is shown in Fig.~\ref{fig12}(c) for
$\alpha-$Tb$_2$CO$_2$ and $\alpha-$Dy$_2$CO$_2$.

\begin{figure}[t]
\includegraphics[width=1.0\columnwidth]{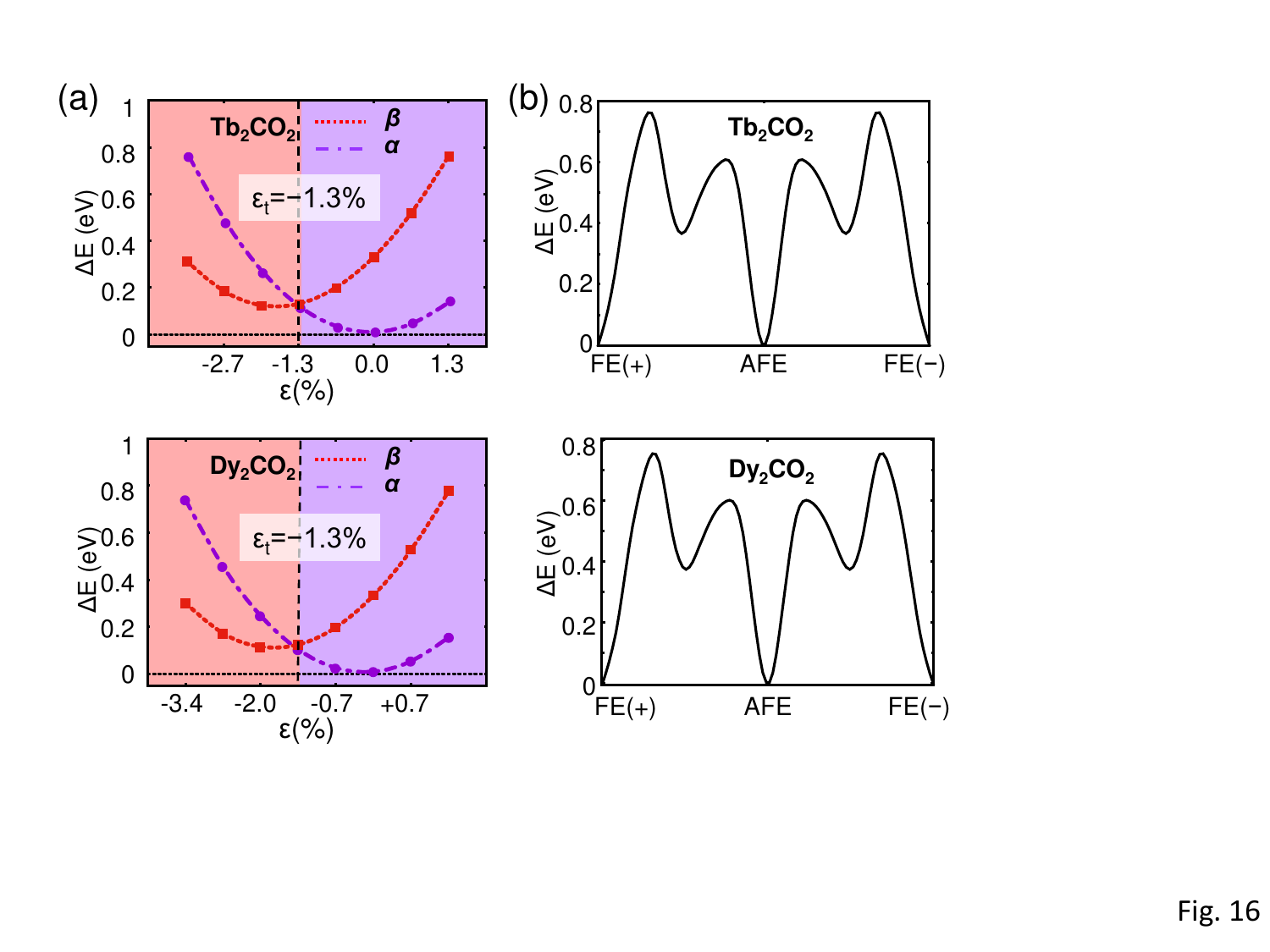}
\caption{%
(a) Energy changes $\Delta{E}$ in the $\alpha$ and $\beta$ phases
of Tb$_2$CO$_2$ and Dy$_2$CO$_2$ as a function of applied in-plane
biaxial strain $\varepsilon$. The region, where the $\alpha-$phase
is more stable, is indicated by the purple background and the
region, where the $\beta-$phase is more stable, by the red
background. Equal stability of both phases at $\varepsilon_{t}$ is
indicated by the dashed line, which separates the two regions. %
(b) Energy changes $\Delta{E}$ along the transformation path from
the $\alpha-$ to the $\beta-$phase of $\alpha-$Tb$_2$CO$_2$ and
$\alpha-$Dy$_2$CO$_2$ at the strain value $\varepsilon_{t}$. %
\label{fig13}}
\end{figure}

\begin{figure}[b]
\includegraphics[width=1.0\columnwidth]{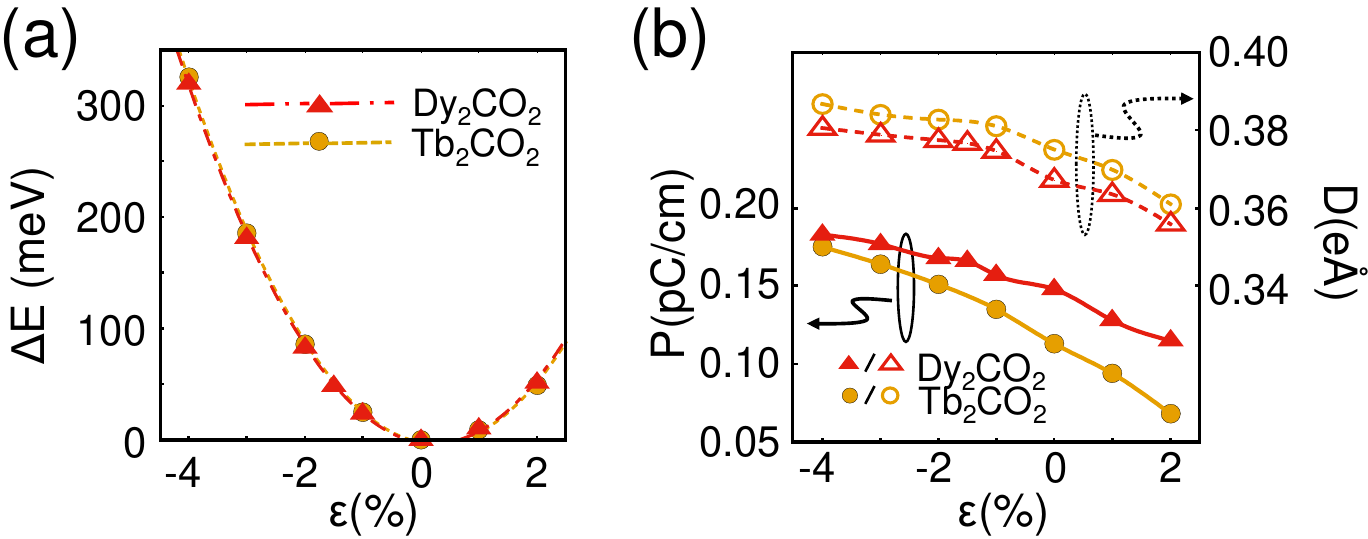}
\caption{%
(a) Energy changes $\Delta{E}$ per unit cell as function of
in-plane biaxial strain $\varepsilon$ applied to %
$\alpha-$Tb$_2$CO$_2$ and %
$\alpha-$Dy$_2$CO$_2$. %
(b) Electric polarization $P$ and dipole moment $D$ of
Tb$_2$CO$_2$ and Dy$_2$CO$_2$ as a function of the in-plane
biaxial strain $\varepsilon$. Results for the two systems are
distinguished by the color and the symbols. %
\label{fig14}}
\end{figure}

\begin{figure}[t]
\includegraphics[width=1.0\columnwidth]{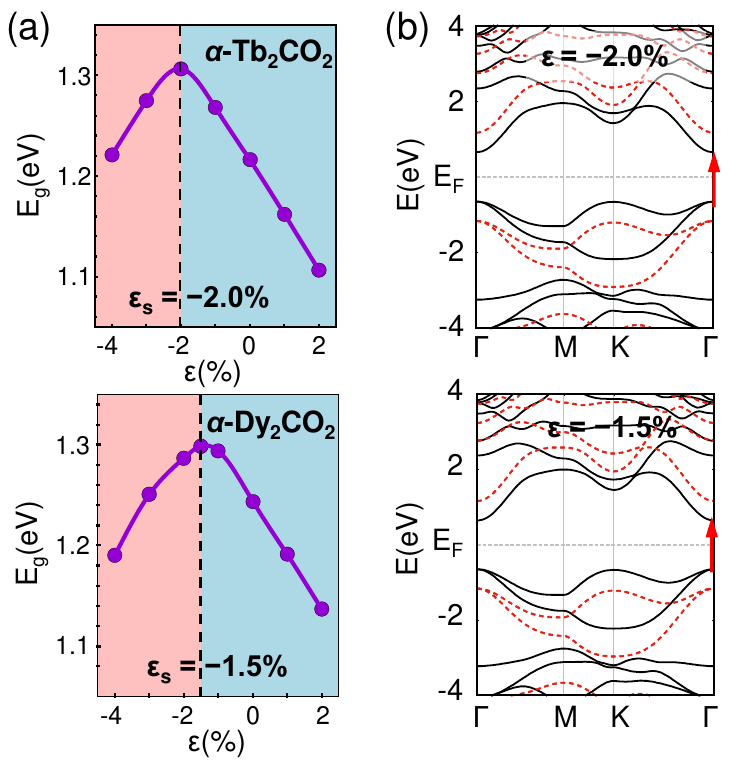}
\caption{%
(a) Dependence of the electronic band gap $E_{g}$ on the in-plane
biaxial strain $\varepsilon$ in $\alpha-$Tb$_2$CO$_2$ and
$\alpha-$Dy$_2$CO$_2$. Both systems change from indirect-gap to
direct-gap semiconductors at compressive strains
$\varepsilon<\varepsilon_{s}$. The direct-gap region at high
compression is indicated by the pink background, and the
indirect-gap region by light blue. %
(b) The electronic bands structure $E(k)$ of $\alpha-$Tb$_2$CO$_2$
and $\alpha-$Dy$_2$CO$_2$ subject to specific strain values
$\varepsilon$. Results based on the DFT-PBE functional are
represented by solid black lines and those based on DFT-HSE06 are
shown by dashed red lines. %
\label{fig15}}
\end{figure}

\begin{figure}[b]
\includegraphics[width=1.0\columnwidth]{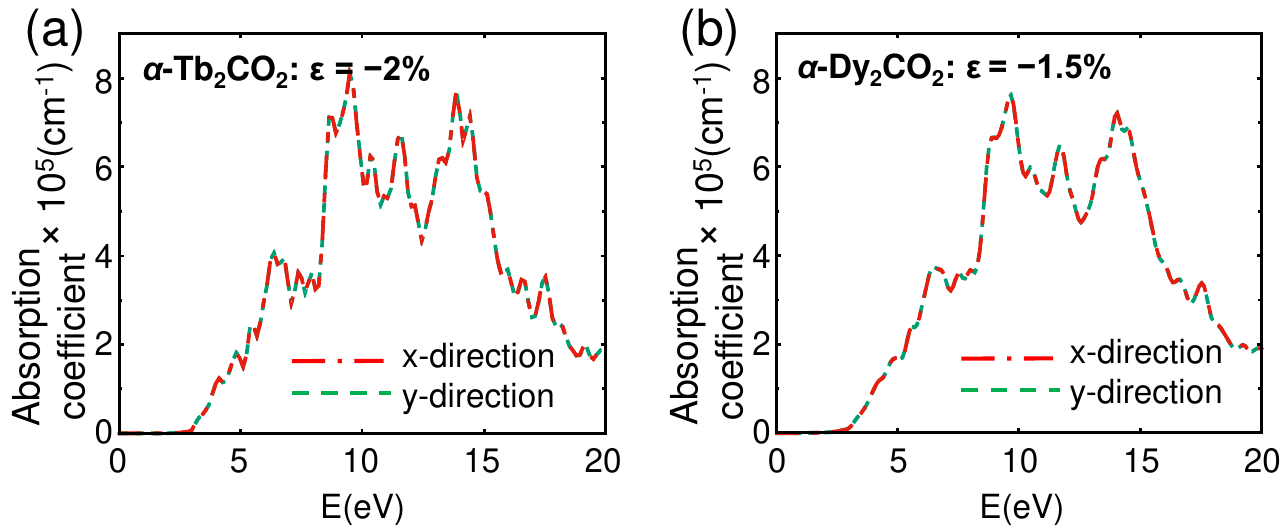}
\caption{%
Optical absorption spectra of %
(a) $\alpha-$Tb$_2$CO$_2$ at $\varepsilon=-2$\% and %
(b) $\alpha-$Dy$_2$CO$_2$ at $\varepsilon=-1.5$\%,
where both systems become direct-gap semiconductors. %
\label{fig16}}
\end{figure}

\subsection{Relative stability of the $\alpha$ and $\beta$
phases of Tb$_2$CO$_2$ and Dy$_2$CO$_2$}

As mentioned in the main manuscript, at strain values below
$\varepsilon_{t}{\approx}-1.3$\% in the compressive regime, the
$\beta-$phase of Gd$_2$CO$_2$ becomes more stable than the
$\alpha-$phase. We see in Fig.~\ref{fig13}(a) that the transition
strain maintains its value $\varepsilon_{t}{\approx}-1.3$\% also
in Tb$_2$CO$_2$ and Dy$_2$CO$_2$. Energy changes along the
transformation path from the $\alpha-$ to the $\beta-$phase, shown
in Fig.~\ref{fig13}(b), indicate a similar energy barrier around
$750$~meV in Tb$_2$CO$_2$ and Dy$_2$CO$_2$ to that found in
Gd$_2$CO$_2$.

\subsection{Strain effect on the electronic structure and polarization
            of Tb$_2$CO$_2$ and Dy$_2$CO$_2$}

Energy changes caused by applying in-plane biaxial strain to
$\alpha-$Tb$_2$CO$_2$ and $\alpha-$Dy$_2$CO$_2$ are shown in
Fig.~\ref{fig14}(a). As seen in Fig.~\ref{fig14}(b) and similar to
$\alpha-$Gd$_2$CO$_2$, the electric polarization and dipole moment
of Tb$_2$CO$_2$ and Dy$_2$CO$_2$ increase with increasing
compressive strain and decrease with increasing tensile strain.
Changing from tensile strain $\varepsilon=+2$\% to compressive
strain $\varepsilon=-4$\% increases the polarization by $257$\% in
$\alpha-$Tb$_2$CO$_2$ and by $159$\% in $\alpha-$Dy$_2$CO$_2$.

Changes in the band gap of $\alpha-$Tb$_2$CO$_2$ and
$\alpha-$Dy$_2$CO$_2$ as a function of biaxial in-plane strain are
shown in Fig.~\ref{fig15}(a). At the compressive strain values
$\varepsilon_{s}=-2.0$\%  for $\alpha-$Tb$_2$CO$_2$ and
$\varepsilon_{s}=-1.5$\% for $\alpha-$Dy$_2$CO$_2$, both systems
turn into direct-gap semiconductors. Their band structure at the
strain value $\varepsilon_{s}$ is shown in Fig.~\ref{fig15}(b).

\begin{figure}[t]
\includegraphics[width=1.0\columnwidth]{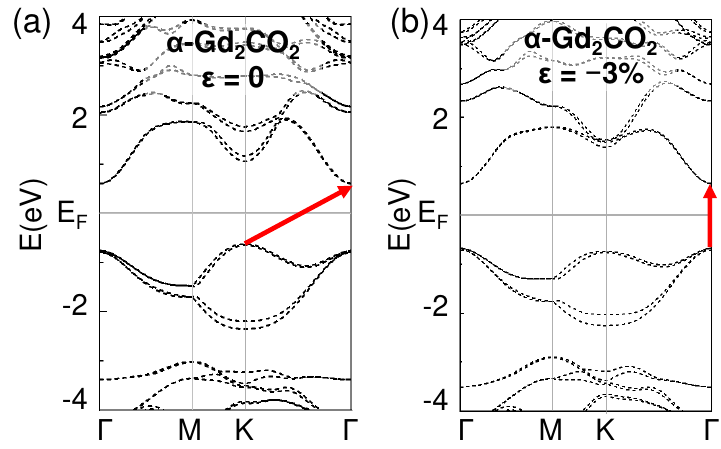}
\caption{%
\modR{%
Electronic band structure $E(k)$ of
$\alpha$-Gd$_2$CO$_2$ subject to %
(a) no strain and %
(b) compressive strain $\varepsilon=-3$\%. The calculations, which
consider spin-orbit coupling (SOC), are based on the DFT-PBE functional. %
}%
\label{fig17}}
\end{figure}

\subsection{Optical absorption of Tb$_2$CO$_2$ and Dy$_2$CO$_2$}

Calculated optical absorption spectra of Tb$_2$CO$_2$ and
Dy$_2$CO$_2$ at the specific biaxial in-plane strain value
$\varepsilon$, where they become direct-gap semiconductors, are
shown in Fig.~\ref{fig16}. Similar to Gd$_2$CO$_2$, the absorption
spectra of Tb$_2$CO$_2$ and Dy$_2$CO$_2$ are characterized by five
absorption peaks in the range  between $5$~eV and $20$~eV. Two of
these peaks are associated with the UV-spectrum of carbon, and the
other three belong to the extreme-ultraviolet spectrum.

\subsection{Effect of spin-orbit coupling on the electronic band
            structure of relaxed and strained Gd$_2$CO$_2$}

\modR{%
We have calculated the effect of spin-orbit coupling (SOC) on the
electronic structure of unstrained and compressed Gd$_2$CO$_2$
using the DFT-PBE functional. As seen in Fig.~\ref{fig17}, SOC
does cause band splitting. However, the splitting is very small at
the VBM and CBM, which remain at the $\Gamma$ point under
compressive strain. Since the electronic structure at the VBM and
CBM dominates the optical properties, we conclude that the
corresponding calculation does not require specific consideration
of SOC effects in M$_2$CO$_2$
monolayers. %
}%


\begin{acknowledgements}
L. Han, P. Liu and D. Liu acknowledges financial support by the
Natural Science Foundation of the Jiangsu Province Grant No.\
BK20210198, the National Natural Science Foundation of China
(NNSFC) Grant No.\ 12204095, the High Level Personnel Project of
Jiangsu Provience Grant No.\ JSSCBS20220120, and the Zhishan
Foundation of Southeast University Grant No.\ 2242023R10006. X.
Lin acknowledges financial support by NNSFC Grant No.\ 11974312.
We dedicate this work to our co-author X.~Lin, who passed away
prematurely while working on this manuscript.
\end{acknowledgements}

%

\end{document}